\documentstyle[epsfig]{article}
\begin{document}
\vspace{3.0cm}
\begin{center}
{\huge \bf Measurements of the reaction {\huge \boldmath 
$\bar{p}p \to \phi\eta$}
of antiproton annihilation at rest at three hydrogen target densities} \\
\vspace{2.0cm}
{\huge The OBELIX Collaboration.}\\ 
\vspace{3.0cm}
{\Large Abstract} \\
\end{center}
\vspace{1.0cm}
{\large The $\bar pp$ annihilation at rest into the $\phi\eta$
final state was measured for three different target densities:
liquid hydrogen, gaseous hydrogen at NTP and at a low pressure  of $5$ mbar.
The yield of this reaction in the liquid hydrogen target is smaller than
in the low-pressure gas target. The branching ratios of the $\phi\eta$
channel were calculated on the basis of 
simultaneous analysis of the three data samples.
The branching ratio for annihilation
into $\phi\eta$ from the $^3S_1$ protonium state turns out to be about
ten times smaller as compared to the one from the $^1P_1$ state.}
\newpage

The Okubo-Zweig-Iizuka (OZI) rule \cite{OZI} allows one to relate  the
ratio between  the cross sections or annihilation  frequencies
of $\phi$ and $\omega$ meson production to the mixing angle
$\Theta$ of the vector meson nonet:
\begin{eqnarray}
R&=&\frac{{\cal F}(\bar{p} + p\to \phi + X)}{{\cal F}
(\bar{p} + p\to \omega + X)} = tan^2 \delta \cdot f =
4.2\cdot 10^{-3}\cdot f
\end{eqnarray}
Here $\delta = \Theta - \Theta_0$, where $\Theta=39^0$
from the quadratic Gell-Mann-Okubo mass formula, 
$\Theta_0=35.3^0$ is the ideal mixing angle and
$f$ is a kinematic phase space factor.
It is known that the predictions of the OZI rule
are fulfilled in the hadronic interactions
at a level of about $10\%$ (for a review, see \cite{Zelenogorsk}).
Recently large OZI rule violation has been found in some $\bar{p}p$
annihilation channels in the experiments with stopped antiprotons at LEAR
(CERN). Measurements of the Crystal Barrel collaboration
show that the ratio of yields for the reactions
$\bar{p} + p \to \phi(\omega) + \gamma $
is $R(\phi \gamma /\omega \gamma)= (250\pm 89)\cdot 10^{-3}$ \cite{OZIgamma}.
Large apparent violation of the OZI rule was also found for the
$ \bar{p} + p \to \phi(\omega) + \pi^0 $ channel,  where
$R(\phi \pi^0 /\omega \pi^0)= (96\pm 15)\cdot 10^{-3}$
for annihilation in the liquid hydrogen target \cite{OZIgamma}
and
$R(\phi \pi^0 /\omega \pi^0)= (114\pm 24)\cdot 10^{-3}$
for annihilation in the gas hydrogen target \cite{Abl.95}.
Conservation of P- and C-parities  allows the
$\bar{p} p \to \phi \pi$ reaction only from the
$^3S_1$ or $^1P_1$ initial states. An interesting dynamical selection
rule was found for this channel: 
$\phi$ production is at least 15 times more
prominent for annihilation from the spin triplet $^3S_1$ initial state
than from the singlet $^1P_1$ one \cite{Pra95}.

On the other hand, there are channels of $\phi$ meson production
in $\bar{p}p$ annihilation at rest where
no OZI rule violation was observed. An example is the reaction
\begin{eqnarray}
\bar{p} + p  & \longrightarrow & \phi + \eta, \label{phieta}
\end{eqnarray}
where the Crystal Barrel collaboration found \cite{CBPhiEta},\cite{OZIgamma}
that the ratio R for annihilation in liquid
$R(\phi \eta/\omega \eta) = (6.0\pm 2.0)\cdot 10^{-3}$  is in
agreement with the OZI rule prediction \cite{OZI}.
Since reaction (\ref{phieta}) proceeds from the same protonium levels
as the $\bar{p} p \to \phi\pi^0$ reaction,
but from isospin I=0, one may expect that the $\phi \eta$ final
state will also be suppressed for annihilation from the $^1P_1$ state.
Surprisingly enough, the result reported here shows that the situation
is opposite: the yield of reaction (\ref{phieta})
for annihilation from the P-wave is higher than from the
S-wave.

We performed systematic measurements of the channel
\begin{eqnarray}
\bar{p} + p  & \longrightarrow & K^+ + K^- + \eta \label{KKeta}
\end{eqnarray}
at rest in a hydrogen target of three densities: 
liquid, gaseous at NTP  and at 5 mbar.
Changing the target pressure allowed selection of the
$\bar{p}p$ initial state. The fraction of annihilation from the 
S-wave decreases with decreasing target density. According to
\cite{Batty}, it varies from $87\%$ to $20\%$
for annihilation in liquid and 5 mbar hydrogen, respectively.

Earlier we reported \cite{Abl.95} measurements of 
reaction (\ref{phieta}) for annihilation in a gas target at NTP,
though with considerably  lower statistics.
The measurements of this reaction for the 5 mbar hydrogen target are made
for the first time.

The experiment was performed at LEAR (CERN), using the
OBELIX spectrometer \cite{Detector}.
The experimental setup consists of detectors arranged
around the Open Axial Field Magnet, which provided the magnetic field
of 0.5 T along the beam axis.
 The Time of Flight (TOF) system
contains two coaxial barrels of plastic scintillators for charged particle
identification and the trigger. The Jet Drift Chamber (JDC) provides
tracking and particle identification by
energy loss measurement. The High Angular Resolution Gamma Detector (HARGD)
is designed to detect neutral
particles by their decay into $\gamma$.
Only the information of the TOF and JDC systems was used in the present
analysis.
All data were collected with the trigger requesting two hits in the inner
barrel of scintillators and two hits in the outer one.

Two-prong events
with the total charge zero and with the length of both tracks more than
20 cm for each data sample were selected.
Kaons were identified by dE/dx measurements. We required both
particles to be identified as kaons.
The number of events in the data samples, the number of events that passed
quality cuts, and particle identification $N_{KKX}$ are presented in Table 1.

\begin{table}[ht]
\caption{Summary of the collected statistics.
The number of events on the tape, which passed quality cuts
(two-prong events with the total charge zero and with the length of 
both tracks more than 20 cm) and were identified as kaons are given.}
\begin{tabular}{lccc} \hline
Target & On tape $(\cdot 10^6)$ & After quality cuts
$(\cdot 10^6)$ & $N_{KKX}~(\cdot 10^5)$ \\ \hline
Liquid        & $10.4$ & $6.7$ & $1.81$ \\
Gas at NTP    & $ 6.7$ & $4.5$ & $1.06$ \\
Gas at 5 mbar & $ 9.4$ & $5.4$ & $1.71$ \\ \hline
\end{tabular}
\end{table}

The experimental distributions of the $K^+ K^- X$ final state for three data
samples without correction for the acceptance
are shown in Fig. 1 (left to right:
liquid, gas at NTP and gas at 5 mbar). 
\begin{figure}
\psfull
\epsfig
{file=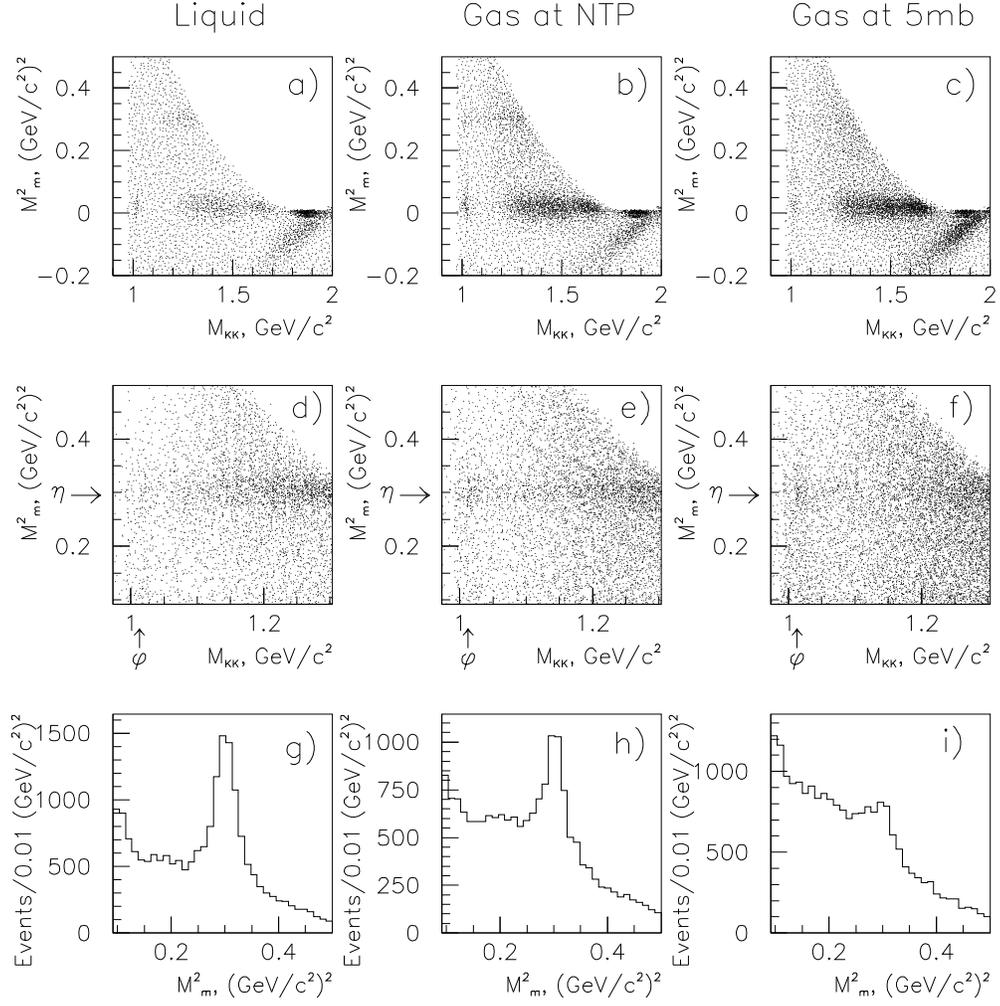,width=15.0cm,height=15.0cm,bbllx=10,bblly=100,bburx=582,bbury=672,clip=}
\caption{Scatter plots of missing masses {\it vs} $M_{K^+K^-}$
(a,b,c). The same distributions around the $\eta$ band are shown in
(d,e,f); missing mass spectra (g,h,i).
The left column corresponds to the data at the liquid target, the middle one 
to the target at NTP and (c,f,i) to the target at 5 mbar.}
\end{figure}
The scatter plots of missing mass squared
distributions {\it vs} $M_{K^+K^-}$ are shown in Figs. 1a-c.
One can see two horizontal bands from reactions with $\pi^0$
and $\eta$. The accumulation of events seen 
on the left side of the plots corresponds to the $\phi \pi^0$ final state. 
In the right corner of the plots there is 
a spot from $\bar{p}p \to K^+K^-$ reaction
and an inclined band from  the $\pi^+ \pi^- X$ final state with wrongly
identified pions.
To see more clearly the region of these plots
around the $\eta$ band we zoomed them in the centre (Fig.1 d,e,f).
A blob from the $\phi\eta$ events is better seen in these plots.
Figures 1g,h,i  show the distributions of the
missing mass square recoiling against the two charged kaons, where 
the peaks due to reaction (\ref{KKeta}) are seen.

To select events from the $\phi\eta$ reaction
the invariant mass distribution of two kaons $M_{K^+K^-}$
for events with the missing mass
around the mass of $\eta$  was analysed. It is shown in the middle part
of Fig.2, where events with
the missing mass interval
$0.26<M^2_{miss}<0.34~~GeV^2/c^4$
(centred around $m^2_{\eta}=0.3~~GeV^2/c^4$) were selected.
The left and right parts of
Fig.2 correspond to the $M_{K^+K^-}$ distributions for
the missing mass intervals below and above the $\eta$ mass, where
number of events with $\eta$ is negligible:
$0.15<M^2_{miss}<0.23$ (left) and
$0.37<M^2_{miss}<0.45~~GeV^2/c^4$ (right).
For the sake of brevity we label the missing mass intervals 
``A", ``B" ,``C" in order of increasing $M^2_{miss}$.
\begin{figure}
\psfull
\epsfig
{file=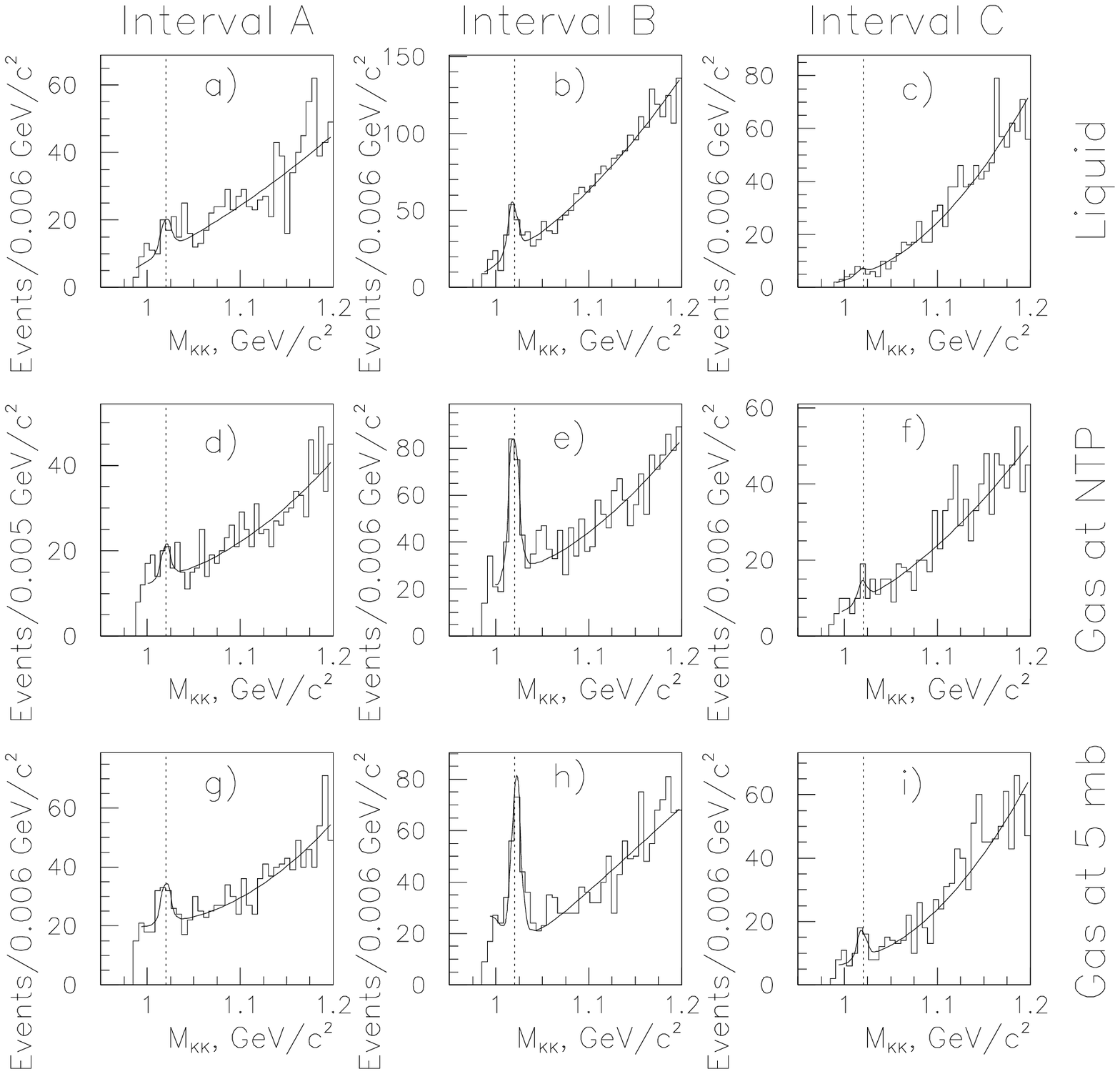,width=15.0cm,height=15.0cm,bbllx=10,bblly=100,bburx=582,bbury=672,clip=}
\caption{The distributions on $M_{K^+K^-}$ for three intervals of missing
mass for three target densities: (a,b,c) for the liquid target, 
(d,e,f) for the gas target at NTP and (g,h,i) for the gas
target at 5 mbar. The left column corresponds to interval ``A"
(see the text), column (b,e,h) corresponds to interval ``B", and column (c,f,i)
corresponds to interval ``C". Dashed lines correspond to 
the $\phi$ meson mass.}
\end{figure}

To illustrate the choice of the intervals in Fig.3 the
distributions on $M^2_{miss}$ obtained by the Monte Carlo simulation
of reaction (\ref{phieta}) and the main background reactions
are shown.
\begin{figure}
\psfull
\epsfig
{file=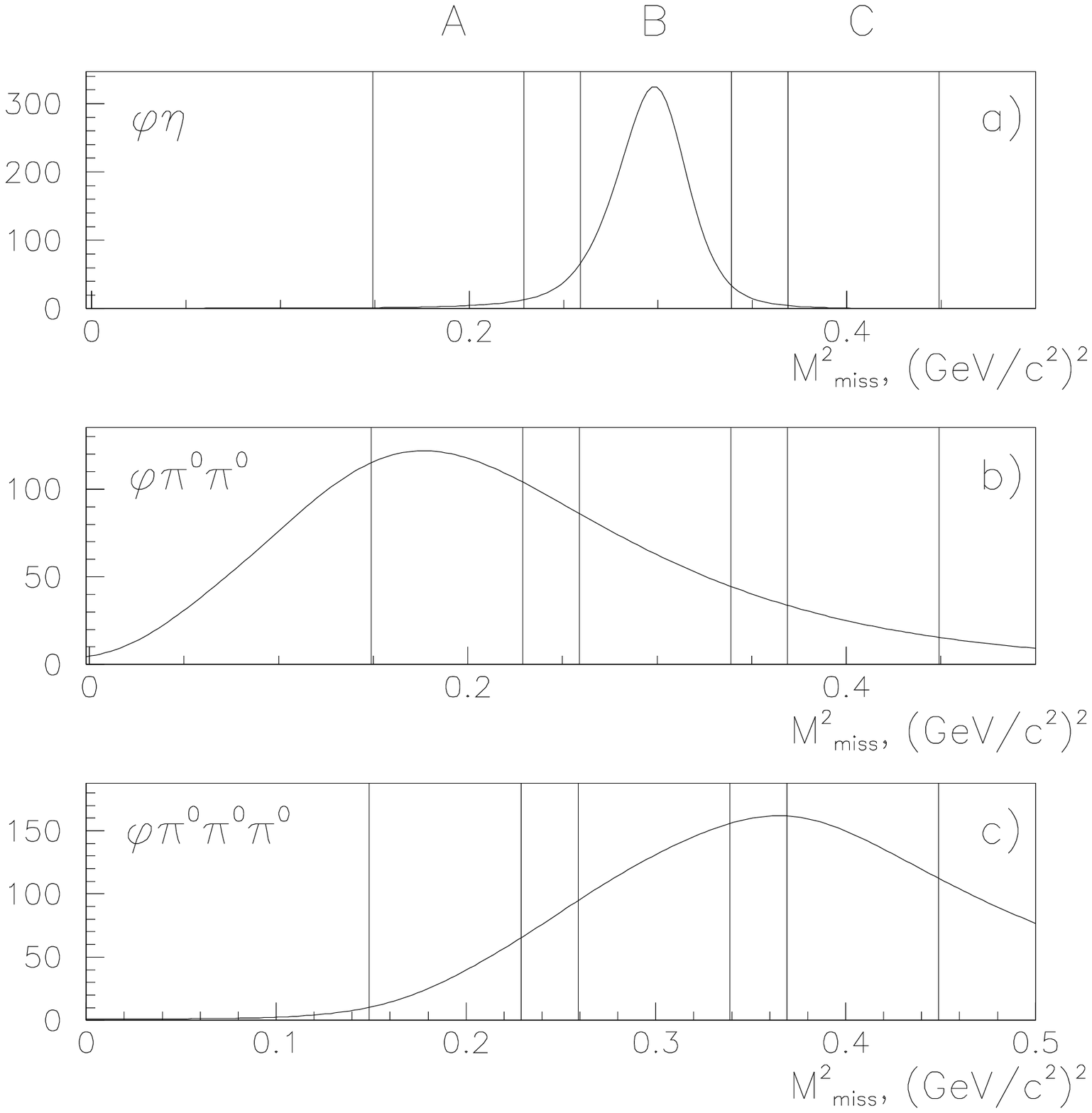,width=9.4cm,height=9.4cm,bbllx=-120,bblly=120,bburx=452,bbury=692,clip=}
\caption{The spectra of missing mass recoiling against $K^+K^-$
system for
a) $\bar{p}p \rightarrow  \phi\eta$ reaction, 
b) $\bar{p}p \rightarrow  \phi\pi^0\pi^0$ and 
c) $\bar{p}p \rightarrow  \phi\pi^0\pi^0\pi^0$,
obtained by Monte Carlo. Lines show the interval limits.}
\end{figure}
One can see that the interval
``B" contains most of the events from the reaction of interest.
The background reaction
\begin{eqnarray}
\bar{p} + p & \longrightarrow & \phi + \pi^0 + \pi^0 \label{phi2pi0}
\end{eqnarray}
gives the main contribution in interval ``A" , whereas the
events from the $\phi \pi^0 \pi^0 \pi^0 $
final state concentrate mainly in the interval ``C".

The experimental distributions for the interval ``B" (Figs. 2b, 2e and 2h) 
for events with missing mass around the mass of $\eta$
show clear peaks in the region of the $\phi$ meson.
The $\phi$ signal for events from 
neighbouring missing mass intervals
``A" and ``C" is far less pronounced. 

To describe the $\phi$ peak in $M_{K^+K^-}$ spectra
we used the Breit-Wigner function, corrected for phase space, 
centrifugal factor and convoluted with a gaussian
experimental resolution function.
The background was treated by a smooth third-order polynomial function.

The $\phi$ peak positions are $1018\pm 1~MeV/c^2,  1019\pm 1~MeV/c^2$ and
$1020\pm 1~MeV/c^2$ for the LQ, NTP and 5 mbar data samples, respectively;
these values agree well with the PDG value for the $\phi$ mass.
The widths of the gaussian, which corresponds to the detector resolution, 
are $4.3\pm 0.3~MeV/c^2, ~3.4\pm 0.3~MeV/c^2$
and $3.7\pm 0.4~MeV/c^2$ for the LQ, NTP and 5 mbar data samples, respectively.
They are in agreement with those obtained by the Monte Carlo simulation.

To evaluate contamination from the background reactions
$\bar{p} p\to \phi X$  in missing mass interval ``B", the number of
events with $\phi$ mesons in neighbouring intervals
``A" and ``C" was determined. The background was subtracted
assuming smooth behaviour of the squared missing mass distribution.
Monte Carlo simulation of different background channels
confirmed this assumption. 

The main background contribution is from the reaction
$ \bar{p} p \to \phi \pi^0 \pi^0 $.
Using the preliminary results for the frequency of this channel obtained by
the Crystal Barrel collaboration with the
liquid target  ${\cal F}(\phi\pi^0\pi^0)=(0.88\pm 0.15)\cdot
10^{-4}$ \cite{CBphi2pi0} one may estimate the
number of $\phi\pi^0\pi^0$ events in interval ``B'':
$N_{ev}={\cal F}\cdot N_{ann}\cdot\varepsilon_{reg} =
(0.88\pm 0.15)\cdot 10^{-4}\cdot
(115.9\cdot 10^{6})\cdot (0.165\pm 0.010)\cdot 10^{-2} = 17\pm 3$.
It is in good agreement with
our estimation of the background, which gives $17\pm 13$ events.

We also verify the contamination from the channels
$\bar{p} p \to \phi X$, where $X$ decays into charged particles
which are not registered by our apparatus. 
For this purpose, the registration efficiency was determined for the reaction
$\bar{p}p\to\phi\rho$, when both pions from $\rho$ decay were missed.
It turns out to be
$\varepsilon_{reg}= 1.0\cdot 10^{-5}$. With the known frequency of
$\bar{p} p \to \phi \rho$ in NTP
${\cal F}=(3.4\pm 0.8)\cdot 10^{-4}$ \cite{PhiRho},
the estimated number of events from this reaction in interval  ``B" is
$N_{ev} \approx 0.2\pm 0.1$.

The annihilation frequency ${\cal F}$ is
\begin{equation}
{\cal F} =\frac{N_{ev}} {N_{ann} \cdot \varepsilon_{reg}}
\end{equation}
where $N_{ann}$ is the  number of annihilations in the target and
$\varepsilon_{reg}$ is the registration efficiency that takes into
account the correction for all $\phi$ and $\eta$ decay modes.

The simulation of reaction (\ref{phieta}) shows that the
registration efficiency $\varepsilon_{reg}$ for annihilation from
the $^1P_1$ initial state is less than from the $^3S_1$ one. 
Without information about relative abundances of these initial
states for annihilation in different targets one
could determine only an upper and lower limits for
the annihilation frequency, corresponding to the initial states
$^1P_1$ and $^3S_1$, respectively.

In Table 2, the number of annihilations $N_{ann}$,
the number of events from the reaction $N_{ev}$,
the registration efficiency $\varepsilon_{reg}$
and the upper and lower limits for the frequencies of the channel 
$\bar{p}p\to\phi\eta$ for each sample are presented.

\begin{table}
\caption{Number of annihilations $N_{ann}$, number of events $N_{ev}$,
registration efficiency $\varepsilon_{reg}$ and upper and lower limits
of the annihilation frequencies $\cal F$ for each sample.}
\begin{tabular}{lccccc} \hline
Target & $N_{ann}\cdot 10^6$ & $N_{ev}$ &
$\varepsilon_{reg},\%$ & ${\cal F} \cdot 10^4$ & Initial state \\ \hline
Liquid &$115.9$ &   $~68\pm 22$ & $0.97\pm 0.02$ & $0.60\pm 0.20$ & $^3S_1$ \\
       &  & & $0.58\pm 0.02$ & $1.01\pm 0.33$
& $^1P_1$ \\  \hline
Gas at NTP & $74.5$ &  $~146\pm 28$ & $1.88\pm 0.03$ & $1.04\pm 0.20$ & $^3S_1$\\
    &   &  & $1.28\pm 0.03$ & $1.53\pm 0.29$
& $^1P_1$ \\ \hline
Gas at 5 mbar & $85.9$ &  $~152\pm 37$ & $1.68\pm 0.03$ & $1.05\pm 0.26$ & $^3S_1$ \\
 &   &  & $1.09\pm 0.03$ & $1.62\pm 0.40$ & $^1P_1$ \\ \hline
\end{tabular}
\end{table}

The systematic uncertainties in the evaluation of the annihilation
frequencies are about  $8\%$. The systematics includes  
beam monitoring, trigger inefficiencies and the uncertainties due to 
varying the $dE/dx$ kaon selection criteria.

It is possible to determine the branching
ratios {\it B.R.} of the reaction $\bar{p}{p}\to \phi\eta$
by using additional information from some models 
of the protonium atoms. Let us divide the number of $\phi$ events $N_{ev}$
in each data sample into two contributions
\begin{eqnarray}
N_{ev}& = & N_{ev}(^3S_1)+N_{ev}(^1P_1) \label{split}
\end{eqnarray}
which are defined, according to Batty \cite{Batty}, as
\begin{eqnarray}
N_{ev}(^3S_1)& = &(1-f_p)\cdot \frac{3}{4}\cdot E(^3S_1)\cdot
B.R.(^3S_1)
\cdot N_{ann}\cdot \varepsilon_{reg}(^3S_1)
\label{3s1} \\
N_{ev}(^1P_1)& = &~~f_p\cdot \frac{3}{12}\cdot E(^1P_1)\cdot
B.R.(^1P_1)
\cdot N_{ann}\cdot \varepsilon_{reg}(^1P_1)
\label{1p1}
\end{eqnarray}

Here $f_p$ is the fraction of annihilations from the P-states,
factors $\frac{3}{4}$ and
$\frac{3}{12}$ are the statistical weights of the $^3S_1$ and
$^1P_1$ states, respectively. $E(^3S_1)$ and $E(^1P_1)$ are the enhancement
factors which reflect deviations from the pure statistical
population of the level. 
The values of $f_p$ and enhancement factors are taken from \cite{Batty}
and shown in Table 3.

The branching ratio B.R. is the probability that the $\bar{p}p$ system
with the definite quantum numbers $J^{PC}$ of the initial state 
annihilates into a given final state. Since our measurements  at three 
hydrogen target densities give three values of $N_{ev}$, it is possible 
to determine the
two unknown parameters $B.R.(^3S_1\to\phi\eta)$ and $B.R.(^1P_1\to\phi\eta)$
from eqs. (\ref{split})-(\ref{1p1}).

By using the enhancement factors from \cite{Batty} for the DR1 model and
considering for the annihilation frequency of
$\bar{p}p \to \pi^0 \pi^0$ reaction in liquid the
value of $(2.8 \pm 0.4)\cdot 10^{-4}$ \cite{Bologna},
we obtain a different set of values for $f_p$ shown in Table 3, which
is useful for testing the stability of our results. 

\begin{table}[ht!]
\caption{Fraction of P-wave annihilation $f_p$ taken
from the 
analyses of \cite{Batty} and \cite{Bologna} and the enhancement factors
E from \cite{Batty}.}
\begin{tabular}{l|c|c|c|c} \hline
& \multicolumn{2}{c|}{$f_p$} & \multicolumn{2}{c}{Enhancement factors}
\\ \cline{2-5}
Target & \cite{Bologna} & \cite{Batty} & $E(^3S_1)$ & $E(^1P_1)$ \\ \hline
Liquid        & $0.06\pm 0.01$ & $0.13\pm 0.04$ & $0.989$ & $0.856$ \\
Gas at NTP    & $0.56\pm 0.04$ & $0.58\pm 0.06$ & $0.993$ & $0.974$ \\
Gas at 5 mbar & $0.84\pm 0.03$ & $0.80\pm 0.06$ & $0.985$ & $0.999$ \\ \hline
\end{tabular}
\end{table}

        The branching ratios obtained from the fit of the events at
three densities are:
\begin{eqnarray}
B.R.(^3S_1\to\phi\eta) & = &(0.76\pm 0.31)\cdot10^{-4} \label{r1} \\
B.R.(^1P_1\to\phi\eta) & = &(7.78\pm 1.65)\cdot10^{-4}
\end{eqnarray}
for the parameters choice
\cite{Batty} and
\begin{eqnarray}
B.R.(^3S_1\to\phi\eta) & = &(0.82\pm 0.28)\cdot10^{-4}  \\
B.R.(^1P_1\to\phi\eta) & = &(7.59\pm 1.55)\cdot10^{-4} \label{r2}
\end{eqnarray}
for the parameters from \cite{Bologna}.
Therefore, for both sets of $f_p$ the branching ratio of
$\bar{p} p \to \phi \eta$ channel for annihilation from
the spin singlet  $^1P_1$ state is about 10 times higher than the
branching ratio from the spin triplet $^3S_1$ state.
An opposite trend was observed for the $\bar{p} p \to \phi \pi^0$
channel \cite{Pra95}.

Using the branching ratios obtained above,
the annihilation frequencies at different hydrogen target densities
can be calculated:

\begin{eqnarray}
{\cal F} & = &(1-f_p)\cdot \frac{3}{4}\cdot E(^3S_1)\cdot
B.R.(^3S_1) +
f_p\cdot \frac{3}{12}\cdot E(^1P_1)\cdot B.R.(^1P_1). \label{f}
\end{eqnarray}

The corresponding results are presented in Table 4.

\begin{table}[ht]
\caption{Annihilation frequencies ${\cal F}\cdot 10^{4}$
calculated from (\ref{f})
with B.R. from (\ref{r1})-(\ref{r2}).
The upper numbers correspond to the P-wave fraction calculated
with the parameters of
\cite{Batty} and the lower numbers correspond to the parameters
of \cite{Bologna}. }
\begin{tabular}{lccc} \hline
Reference & Liquid & Gas at NTP & Gas at 5 mbar \\ \hline
\cite{Batty}   & $0.71\pm 0.07$ & $1.33\pm 0.15$ & $1.66\pm 0.20$ \\
\cite{Bologna} & $0.67\pm 0.07$ & $1.30\pm 0.14$ & $1.69\pm 0.21$ \\
\hline
\end{tabular}
\end{table}

One could see that in spite of the significant difference in the branching
ratios from the S- and P-wave states
the annihilation frequencies increase only by a factor of
two from the liquid to the 5 mbar sample. It is due to the relatively
small statistical weight of the $^1P_1$ state.

Several experimental studies of the $\phi\eta$
antiproton annihilation channel at rest have been reported.
The results
of these measurements are shown in Table 5.
The Crystal Barrel collaboration measured the $\bar{p}p\to\phi\eta$
channel for annihilation in liquid hydrogen \cite{CBPhiEta}.  The
absolute value for the annihilation frequency was not determined but
the ratio to the $\bar{p}p\to\phi\pi^0$ frequency was given. The value
in Table 5 was deduced from this ratio and from 
the annihilation frequency of the
$\bar{p}p\to\phi\pi^0$ channel measured by the same collaboration
\cite{OZIgamma}.  It is in good agreement with our result for annihilation
in liquid shown in Table 2 and Table 4.

\begin{table}[ht]
\caption{Experimental results of the previous measurements
of the $\bar{p} p \rightarrow \phi \eta$ channel at rest.} 
\begin{tabular}{ccc} \hline
Reference & ${\cal F}\times 10^{4}$ & Comments \\ \hline
~\cite{CBPhiEta}, \cite{OZIgamma} & $0.66\pm 0.19$ &
liquid target \\
\cite{ASTPhiEta} & $0.37\pm 0.09$ & gas at NTP \\
\cite{ASTPhiEta} & $0.41\pm 0.16$ & gas at NTP, LX-trigger \\
\cite{Abl.95} & $1.04\pm 0.33\pm 0.05$ & gas at NTP, upper limit \\
\cite{Abl.95} & $0.74\pm 0.22\pm 0.03$ & gas at NTP, lower limit \\
\hline
\end{tabular}
\end{table}

The ASTERIX collaboration \cite{ASTPhiEta}
measured the $\phi \eta$ channel for antiproton annihilation
in a gas target at NTP. Beside it the yield of (\ref{phieta})
was measured with the so-called LX-trigger, which enriched the data
sample with P-wave annihilation events.
Our results for the gas target at NTP shown in
Table 2 and Table 4
disagree with the absolute values of the ASTERIX collaboration.

The previous measurements of the OBELIX collaboration \cite{Abl.95},
performed with a gas NTP target with another data set and 
with a lower statistics 
than in the present paper, are in agreement with the results shown in Table 2.

     The main distinctive feature of the $\bar{p} p \to \phi \eta$ channel
is the increase in the annihilation frequency
for annihilation in gas at low pressure. In terms of the branching ratios
it leads to the conclusion that the branching ratio from
the spin singlet  $^1P_1$ state is 10 times higher than the
branching ratio from the spin triplet $^3S_1$ state (see Eqs.
(\ref{r1} - \ref{r2})).
Just the opposite tendency was observed for the $\bar{p} p \to \phi \pi^0$
channel \cite{Pra95}, where the branching ratio from
the $^3S_1$ initial state is at least 15 times larger than 
from the $^1P_1$ one.

The unusual spin dependence of the $\phi \pi^0$ production
was explained in \cite{Ell.95} under the assumption that the
$s \bar s$ pairs in the nucleon are
polarized,
as suggested by the experiments on deep inelastic lepton scattering
(for review of experimental results, see \cite{SMC.97}).
Following this model, the observed
strong violation of the OZI rule is only apparent, because it is due to
additional production of $\phi$ by the connected diagrams that are allowed
if the nucleon wave function contains a number of $s \bar s$
pairs.  The model predicts that the $\phi$ production should be enhanced
from the spin triplet initial states, whereas production of the $\bar{s}s$
pair with total spin $S=0$ should be enhanced from the spin singlet initial
states. Since $\eta$ meson has a substantial $\bar{s}s$ component, the
production of the $\phi \eta$ final state could be regarded as a
production of two $\bar{s}s$ pairs, one in the spin triplet state
and the other in the spin singlet state. The extension of the polarized
intrinsic strangeness model for such case is not straightforward.
We note also that the same
strong enhancement of $\eta$ production from the initial spin singlet
state was observed in 
$pp\to pp\eta$ and $pn\to pn\eta$ reactions \cite{Chi.94}. An attempt
to interpret this effect in the polarized nucleon strangeness model
was done in \cite{Rek.97}.

        In some models opulent production of the $\phi$ meson in
$\bar{p}p$ annihilation at rest is assumed to be due to
the rescattering diagrams with $K^*K$ or
$\rho \rho$ in the intermediate state (see \cite{Gor.96}, \cite{Lev.94})
and references therein). 
Calculations of the triangle diagram with $K^*K$ intermediate states
\cite{Lev.94} give for the branching ratio of the $\phi \eta$ from
$^3S_1$ initial state $B.R.(^3S_1\to\phi\eta)  = (0.3\pm 0.1)\cdot10^{-4}$,
a value not far from our experimental results (\ref{r1}-\ref{r2}). 
However, these approaches could predict neither
the observed spin dependence of the $\phi \pi^0$ final state nor 
the spin dependence of the $\phi \eta$ channel.

An interesting possibility considered in \cite{Mar.97} is that the
final state interaction (FSI) of two kaons could enhance the 
$\phi$ production.
Indeed, for annihilation at rest the space volume of $K K \eta$ final
state is quite limited, two kaons are created with low relative momenta
and could, in principle, fuse into $\phi$ due to the FSI. However 
this model does not explain why the FSI effects are stronger for 
annihilation from the P-wave than from the S-wave.

        An attempt to calculate the $\phi$ yields in $\bar{p}p$
annihilation at rest in a non-relativistic quark model with the $\bar{s}s$
admixture in the nucleon wave function was done in \cite{Gut.97}. 
It was assumed that the $\phi$ is produced as a "shake-out" of the
nucleon $\bar{s}s$ component. The calculated branching ratios 
of the $\phi \eta$ channel are
$B.R.(^3S_1\to\phi\eta)  = (1.4-1.8)\cdot10^{-4}$ and 
$B.R.(^1P_1\to\phi\eta)  = (0.15-0.2)\cdot10^{-4}$. Therefore this
"shake-out" mechanism predicts a trend which is exactly opposite
to that experimentally observed by us.  

The increasing of the $\phi \eta$ yield with the decreasing of the
target density arises the question of the corresponding behaviour
of the $\bar{p} p  \to \omega \eta$ channel. 
This yield was measured only
for the annihilation in liquid \cite{OZIgamma} and it is quite high:
${\cal F}(\omega \eta) = (1.51\pm0.12)\cdot10^{-2}$. If the arguments
of the polarised strangeness model are valid for the $\eta$ production, then
the yield of the
$Y(\omega \eta)$ final state from the $^1P_1$ state should be higher than
from the $^3S_1$ final state.
It would be interesting to verify this prediction experimentally since
we have demonstrated that the branching ratio of the $\phi \eta$
channel significantly increases for the P-wave annihilation. If the
ratio between the $\phi \eta$ and $\omega \eta$ channels does not change 
with the hydrogen target density,
the apparent OZI violation will be absent for annihilation in low pressure
gas as it is for annihilation in liquid. 

In conclusion, the $\bar pp$ annihilation at rest into $\phi\eta$
final state was measured at three different target densities: liquid,
NTP, 5 mbar.
The yield of this reaction for the liquid hydrogen target is smaller than
for the low pressure gas target. The branching ratios of $\phi\eta$
channel were calculated on the basis of 
simultaneous analysis of the three data samples.
The branching ratio for annihilation
into $\phi\eta$ from the $^3S_1$ protonium state turns out to be about
ten times smaller as compared
to the one from the $^1P_1$ state. An opposite trend was observed 
\cite{Pra95} for the $\phi \pi$ final state.

We would like to thank the technical staff of the LEAR machine group for
their support during the runs. We are grateful to J.Ellis,
D.Kharzeev and V.Markushin for the stimulating discussions.


\begin{thebibliography}{999}
\bibitem{OZI}
S. Okubo, Phys. Lett. {\bf B 5} (1963) 165; \\
G. Zweig, CERN Report {\bf No.8419/TH412} (1964); \\
I. Iizuka, Prog. Theor. Phys. Suppl. 37 {\bf 38} (1966) 21.
\bibitem{Zelenogorsk}
M.G. Sapozhnikov, JINR preprint {\bf E15-95-544}, (1995).
\bibitem{OZIgamma}
C. Amsler et al.,Phys. Lett.. {\bf B 346} (1995) 363.
\bibitem{Abl.95}
V.G. Ableev et al., Nucl. Phys. {\bf A 594} (1995) 375.
\bibitem{Pra95}
The OBELIX collaboration. A. Bertin et al., {\em Proceedings HADRON'95
Conference\/}, (Manchester, 1995) 337.
\bibitem{CBPhiEta}
C. Amsler et al. Phys. Lett. {\bf B 319} (1993) 373.
\bibitem{Batty}
C.J. Batty, Nucl. Phys. {\bf A 601} (1996) 425.
\bibitem{Detector}
A. Adamo et al., Sov. J. Nucl. Phys. {\bf 55} (1992) 1732.
\bibitem{CBphi2pi0}
Spanier, S.
Workshop on {\it The Strange Structure of the Nucleon}, CERN, 1997
(unpublished).
\bibitem{PhiRho}
J. Reifenrother, {\em Proceedings LEAP-90 Conference\/}, (Stockholm, 1990).
\bibitem{Bologna}
A. Zoccoli, Proceedings of the HADRON
Conference, Brookhaven (1997), in press.
\bibitem{ASTPhiEta}
J. Reifenrother et al., Phys. Lett. {\bf B 267} (1991) 299.
\bibitem{Ell.95}
J. Ellis et al., Phys. Lett. {\bf B 353} (1995) 319.
\bibitem{SMC.97}
The SMC Collaboration, B. Adeva et al., {\it Phys.Lett.} {\bf B412 }
414 (1997).
\bibitem{Chi.94}
E. Chiavassa et al., Phys. Lett. {\bf B 337} (1994) 192.
\bibitem{Rek.97}
M. Rekalo, J. Arvieux, E. Tomasi-Gustafsson, Phys. Rev. {\bf C 55}
(1997) 2630.
\bibitem{Gor.96}
O.E. Gortchakov, M.P.Locher, V.E.Markushin, S. von Rotz,
Z.Phys. {\bf A 353} (1996) 447.
\bibitem{Lev.94}
D.Buzatu, F.M.Lev , Phys. Lett. {\bf B 329} (1994) 143.
\bibitem{Mar.97}
V.E.Markushin, M.P.Locher, Eur.Phys. J. A1 (1998) 91.
\bibitem{Gut.97}
T.Gutsche, A.Faessler, G.D.Yen, S.N.Yang, Nucl. Phys. 
{\bf B (Proc.Suppl.) 56A} (1997) 311.
\end{thebibliography}
\end{document}